\documentclass[%
 reprint,
 amsmath,amssymb,
 aps,
pra,
]{revtex4-2}

\usepackage{graphicx}
\usepackage{dcolumn}
\usepackage{bm}
\usepackage{hyperref}
\usepackage{dsfont}
\usepackage{xcolor}
\usepackage{float}

\begin{document}

\preprint{APS/123-QED}

\title{Improved Fermionic Scattering for the NISQ Era}

\author{Michael Hite}
 \affiliation{University of Arizona}
 \email{michaelhite@arizona.edu}

\date{\today}

\begin{abstract}
In the era of noisy intermediate scale quantum (NISQ) hardware, digital quantum computers are limited to shallow circuits on the order of a thousand layers due to system noise and qubit decoherence. Thus, every step of a simulation must be as efficient as possible. Modifying the recent Givens Rotation state preparation by Chai et al and ladder operator block encoding method by Simon et al, we propose a scattering state preparation method that approximates the fermionic wave packets by localizing them in space, reducing circuit depth by nearly half, while also preserving fermionic anti-commutation relations. Using MPS simulations, we show that these approximated wave packets approach the exact wave packets in weakly interacting critical theories; and then show its immediate application on modern day hardware with IonQ's Forte 1 machine.
\end{abstract}

\maketitle

\textit{Introduction} - Quantum computers provide a natural pathway for large real-time scattering simulations, but in the noisy intermediate scale quantum (NISQ) era \cite{Preskill2018quantumcomputingin} system noise and qubit decoherence severely limit both simulation time and complexity. In terms of quantum circuit depth on digital hardware, this means we can only have circuit depths of around a thousand gate layers with sophisticated error correction and mitigation techniques. Nevertheless, there have already been very promising results obtained for small volume scattering simulations \cite{Turco_2024,kreshchuk2023simulatingscatteringcompositeparticles,briceño2023coherentquantumcomputationscattering,Sharma_2024,Wang_2024,bennewitz2024simulatingmesonscatteringspin,turro2024evaluationphaseshiftsnonrelativistic,yusf2024elasticscatteringquantumcomputer,qsimscatt,OG-PhaseShiftPaper,parks2023applyingnoxerrormitigation}. In the context of fermionic scattering, Jordan, Lee, and Preskill proposed a six step simulation process where one (1) prepares the ground state of the free theory, adiabatically turns on the (2) nearest neighbor and (3) interacting terms of the Hamiltonian to evolve into the ground state of the interacting theory, (4) excite wave packets, then (5) real-time evolve the state, and finally (6) measurement \cite{jordan2014quantumalgorithmsfermionicquantum, Jordan_2012}. Informed variational algorithms like the variational quantum eigensolver (VQE) \citep{Peruzzo2014} or the improved adapt-VQE allows the state preparation steps (1-4) to be combined and have proved been promising for these calculations \cite{farrell2024quantumsimulationshadrondynamics,farrell2025digitalquantumsimulationsscattering,Zemlevskiy:2024vxt}. Similarly, Davoudi et. al \cite{davoudi2024scatteringwavepacketshadrons} used ancillary qubits to place the non-unitary creation operators in a larger unitary space to make wave packets for hadronic theories.

Suppose we let VQE handle the vacuum preparation (steps 1-3), and use alternative methods to prepare fermionic wave packets. These wave packets are a sum of highly non-local non-unitary fermionic creation operators. For standard state decomposition procedures like M\"ott\"onen\cite{mottonen2004transformationquantumstatesusing} and quantum Shannon\cite{Shende_2006}, circuit depth grows far too quickly per system size to be useful. Thus, specific and efficient state preparation techniques are needed in the near term. A technique that has shown great promise uses Givens rotation matrices \cite{Arrazola_2021,Arrazola_2022,PhysRevA.98.022322}, decomposing the state into a product of approximate particle conserving rotation operators when acting on a non-trivial ground state. Chai et. al \cite{chai2024fermionicwavepacketscattering} have used this to simulate fermion/anti-fermion scattering in the Thirring Model.

We introduce a modification to Chai et. al's method for weakly interacting systems that reduces circuit depth by almost a half via localizing the wave packets in position space. As well, we employ unitary particle creation operator via ladder operator block encoding \cite{Simon2025ladderoperatorblock} to allow for preparation of critical theories.

The letter is structured as follows. We first introduce a toy interacting fermionic Hamiltonian using the transverse field Ising model. We then describe the Givens state preparation method with added approximation and ladder operator block encoding. Then, using matrix product state (MPS) methods, we show that our wave packets are an excellent approximation in the weakly interacting regime. Finally, we implement our method on IonQ's trapped ion Forte 1 computer \cite{ionq_docs}.

\textit{Theory} - Consider a modified $1+1D$ $N$-site Transverse Field Ising model whose Hamiltonian is given by
\begin{equation}\label{eq:ham}
\hat{H}=-\sum_{j=1}^{N}\left(J\hat{\sigma}_{j}^{x}\hat{\sigma}_{j+1}^{x} + h\hat{\sigma}_{j}^{z} + g\hat{\sigma}_{j}^{z}\hat{\sigma}_{j+1}^{z}\right),
\end{equation}
where $J(g)$ is nearest neighbor coupling in the $X(Z)$ direction, $h$ is the transverse field coupling, and $\hat{\sigma}_{N+1}=\hat{\sigma}_{1}$ for periodic boundary conditions. For $g=0$, the model is diagonalized via three transformations: Jordan-Wigner, Fourier, and Bogoliubov \citep{SML,Sachdev,Kaufman,Cervera_Lierta_2018}. The Jordan-Wigner fermionic operators are defined as:
\begin{equation}
\hat{c}_{j}^{\dagger}=\left(\prod_{i=1}^{j-1}-\hat{\sigma}_{j}^{z}\right)\hat{\sigma}_{j}^{-},\qquad\hat{c}_{j}=\left(\prod_{i=1}^{j-1}-\hat{\sigma}_{j}^{z}\right)\hat{\sigma}_{j}^{+},
\end{equation}
Due to the boundary term, the Hamiltonian is dependent upon the particle number of the state, so we define two sets of momentum numbers for even and odd numbers of particles:
\begin{equation}
k=\begin{cases}
0,\pm\frac{2\pi}{N},\pm\frac{4\pi}{N},\dots,\pi & \text{(odd particle number)}\\
\pm\frac{\pi}{N},\pm\frac{3\pi}{N},\dots,\pm\frac{(N-1)\pi}{N} & \text{(even particle number)}
\end{cases}.
\end{equation}
For all $J>0$, the ground state is non-trivial, with a phase transition occurring at $J=h$ in the continuum.

\begin{figure}
    \centering
    \includegraphics[width=0.9\columnwidth]{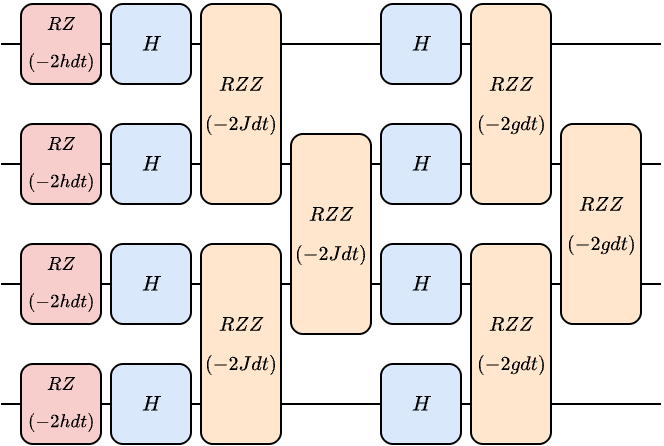}
    \caption{The Trotterized time evolution operator for the modified Ising model. The boundary term is assumed but not shown. All Pauli-rotation gates are defined as $RP(\theta)=\exp(-i\hat{P}\;\theta/2)$. The layers $H$RZZ$H$ correspond to RXX layers, where $H$ is the Hadamard gate.}
    \label{fig:ising-teo}
\end{figure}

The inspiration for the $ZZ$-term comes from the four fermion interaction term in discrete space given by \citep{chai2024fermionicwavepacketscattering}
\begin{align}
4g\sum_{j}\hat{c}_{j}^{\dagger}\hat{c}_{j}\hat{c}_{j+1}^{\dagger}\hat{c}_{j+1} & =g\sum_{j}\left(\hat{\mathds{1}}-\hat{\sigma}_{j}^{z}\right)\left(\hat{\mathds{1}}-\hat{\sigma}_{j+1}^{z}\right)\\
 & =g\sum_{j}\left(\hat{\mathds{1}}-2\hat{\sigma}_{j}^{z}+\hat{\sigma}_{j}^{z}\hat{\sigma}_{j+1}^{z}\right)\nonumber 
\end{align}
The first term is the identity and can be ignored, the second term is the transverse field coupling of the Ising model up to a negative sign, and the third term is what was added in Equation \ref{eq:ham}. Using the Lie-Trotter formula \cite{lie-trotter, suzuki-trotter}, the circuit for time evolution operator (TEO) to first order in $\Delta t$ is given in Figure \ref{fig:ising-teo}. We can go to higher levels of precision with the TEO, but that is not the focus of this letter.

\textit{State Preparation via Givens Rotations} - In position space, a simple fermionic Gaussian wave packet with average position $x_{A}$ and average momentum $k_{A}$ is given by
\begin{equation}
\left|\psi\right\rangle =\mathcal{G}_{A}\left|\Omega\right\rangle =\left(\sum_{j=1}^{N}e^{-ik_{A}j}e^{-\left|j-j_{A}\right|^{2}/\sigma^{2}}\hat{c}_{j}^{\dagger}\right)\left|\Omega\right\rangle ,\label{eq:wave-packet}
\end{equation}
where $\sigma$ is proportional to the width of the wave packet. A two-packet scattering state with position (momentum) centers $x_{A}(k_{A}),\;x_{B}(k_{B})$ is then $\mathcal{G}_{A}\mathcal{G}_{B}\left|\Omega\right\rangle $. The net operator $\mathcal{G}_{A}\mathcal{G}_{B}$ has $N(N-1)/2$ terms. Suppose we truncate the sums on the right moving and left moving wave packets to act on the left and right halves of the space respectively. Then $\mathcal{G}_A\mathcal{G}_B$ now only has $N(N/2-1)/4$ terms. For strongly interacting theories ($g\rightarrow h/2$), the truncation effects cannot be ignored, but in the perturbative weakly interacting sector ($g<<h/2$) the truncation effects are suppressed. The inspiration for the truncation will follow.

\begin{figure*}
    \centering
    \includegraphics[width=0.95\linewidth]{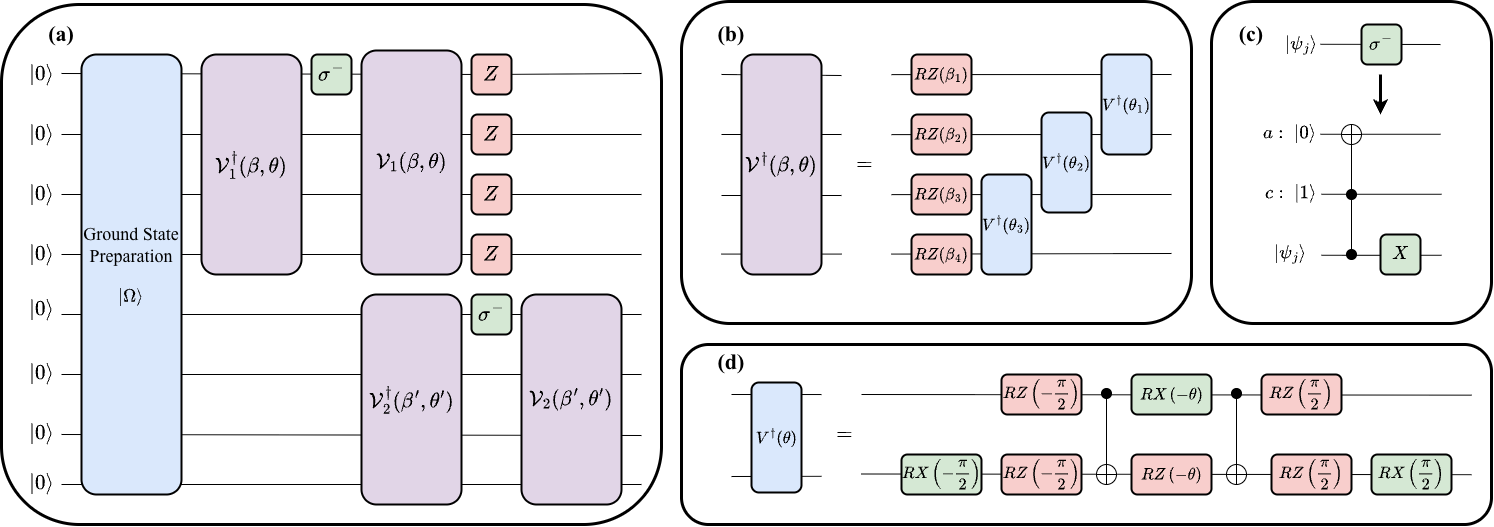}
    \caption{(a) Full scattering state preparation circuit in an 8-site space. The ground state is prepared separately with VQE, then two narrow wave packets are built in subspaces 0-3 and 4-7. The string of Pauli-$z$ gates restore Fermionic anti-commutation relations for the second wave packet. (b) Decomposition of $\mathcal{V}^\dagger(\beta,\theta)$ in terms of single and two qubit operations. (c) Equivalent unitary operation for $\sigma^-$ using a control and ancilla qubit. (d) Decomposition of the two qubit Givens rotation operation in (b). The CNOT depth of the circuit in (a) (excluding ground state preparation) is $18$.}
    \label{fig:Givens-rotation-circuit}
\end{figure*}
We provide a brief overview of the Givens Rotations method of fermionic wave packet preparation on quantum computers (for a full treatment, see \citep{chai2024fermionicwavepacketscattering}). Let $a_{j}=\exp\left(-\left|j-j_{A}\right|^{2}/\sigma^{2}\right)$ and $\beta_{j}=k_{A}j$. The method involves diagonalizing $\left|\psi\right\rangle $ in terms of fermionic operators: 
\begin{equation}
V\hat{c}_{1}^{\dagger}V^{\dagger}=\sum_{j}a_{j}e^{-i\beta_{j}}\hat{c}_{j}^{\dagger}
\end{equation}
by first canceling the phase $\exp\left(-i\beta_{j}\right)$ and then the coefficients $a_{j}$ of each term. The phases are canceled by single qubit $z$-rotations:
\begin{equation}
V^{\dagger}(\beta)=\prod_{j}\text{RZ}\left(\beta_{j}\right)=\exp\left(-\frac{i}{2}\sum_{j}\beta_{j}\hat{\sigma}_{j}^{z}\right).
\end{equation}
The real coefficients are canceled by iteratively diagonalizing each row of $\left|\psi\right\rangle $ with the operators
\begin{equation}
V^{\dagger}\left(\theta_{j}\right)=\exp\left[i\frac{\theta_{j}}{2}\left(\hat{\sigma}_{j-1}^{x}\hat{\sigma}_{j}^{y}-\hat{\sigma}_{j-1}^{y}\hat{\sigma}_{j}^{x}\right)\right],
\end{equation}
acting on nearest neighbors $\left(N-1,N\right),$ $\left(N-2,N-1\right),$ $\dots,\left(1,2\right)$ (no boundary term), for $\theta_{j}=\arctan\left(-a_{j}/a_{j-1}\right)$. Beginning at $\left(N,N-1\right)$, the coefficients ``up'' the vector $a_{j-1}$ are updated after each $V^{\dagger}\left(\theta_{j}\right)$ is applied. The angles are determined classically by diagonalizing in the subspace spanned by the $N$ single particle states. In qubit space, the operator is efficiently decomposed in terms of two CNOT gates and $x$ and $z$ rotations (Figure \ref{fig:Givens-rotation-circuit} (c)). The wave packet creation operator is then
\begin{equation}
\mathcal{G}_{A}=V(\beta)V(\theta)\hat{\sigma}_{1}^{-}V^{\dagger}(\theta)V^{\dagger}(\beta),
\end{equation}
for
\begin{equation}
V^{\dagger}(\theta)=V^{\dagger}\left(\theta_{1}\right)\dots V^{\dagger}\left(\theta_{N-2}\right)V^{\dagger}\left(\theta_{N-1}\right).
\end{equation}
The circuit for two wave packets in an eight-site model is given in Figure \ref{fig:Givens-rotation-circuit} (a). Notice that the addition of the string of $\sigma^z$'s makes the fermionic operator $\hat{c}^\dagger_4$. For a pair of sites the CNOT depth is $4$, which means the $N/2$-site subspace spanned by a single wave packet has CNOT depth $4(N/2-1)$, and the two wave packets can be built in parallel. In the original formulation for fermion-fermion scattering, the CNOT depth would be $8(N-1)$ and the two wave packets are built in series. Thus, in terms of circuit depth, the truncation is advantageous for the NISQ era.

Now we deal with the non-unitary lowering operator $\sigma^-$. Simon et al \cite{Simon2025ladderoperatorblock} introduced the Ladder Operator Block Encoding (LOBE) method where a fermionic creation operator can be placed within a larger unitary $U$ by using a single external control qubit $c$ and an ancilla qubit $a$. Suppose we apply $\sigma^-$ to some qubit $j$ whose in the state $|\psi_j\rangle=a|0\rangle+b|1\rangle$. The resultant state is $\sigma^-|\psi_j\rangle=a|1\rangle$. The operator in Figure \ref{fig:Givens-rotation-circuit} (c) moves the $U b|1\rangle=b|0\rangle$ into an orthogonal state, so the system should only see state $a|1\rangle$, and is confirmed by measuring the ancilla in the $|0\rangle$ state. The Toffoli gate can be decomposed in terms of six CNOT gates \cite{shende2008cnotcosttoffoligates}. Thus, we have a non-variational method to excite truncated wave packets on a ground state that obey fermionic anti-commutation relations whose CNOT depth $4(N/2-1)+6$. We can truncate the regions further ($M<N/2$), an that will be explored in a future work.
\begin{figure}
    \centering
    \includegraphics[width=0.9\linewidth]{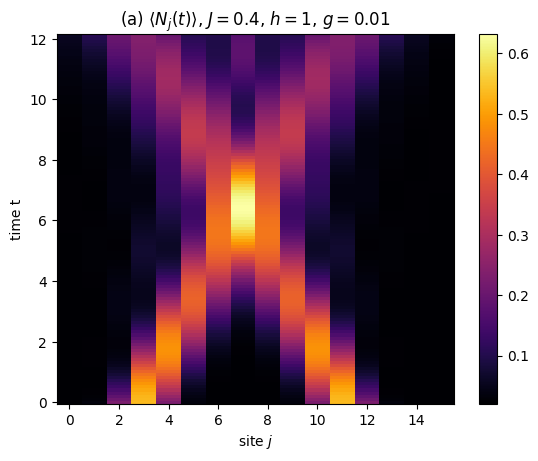}
    \includegraphics[width=0.9\linewidth]{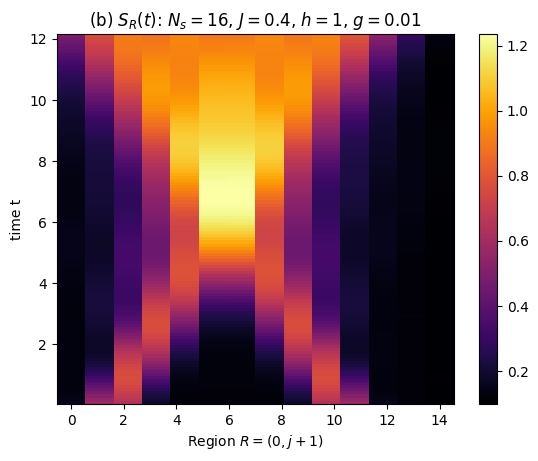}
    \caption{MPS Evolution of (a) site occupations $\langle N_j(t)\rangle$ and (b) bipartite entanglement entropy of regions $(0,j+1)$ to the rest of the system $(j+2,N-2)$ for a scattering state with initial mean positions $x_A=3,\;x_B=11$, momenta $k=\pm7\pi/16$, and width $\sigma=3/2$ for $J=0.4,\;h=1,\;g=0.01,\;\Delta t=0.1$.}
    \label{fig:MPS-example}
\end{figure}
\textit{Results} - We first verify with MPS simulations that our wave packets are a good approximation to the full wave packets in the perturbatively interacting regime. All MPS simulations are done with the ITensor library \cite{ITensorCode, ITensorArticle}. Time evolution is accomplished with time evolving block decimation (TEBD) \cite{TEBD-OG-Paper} and the ground state is prepared with the density matrix renormalization group (DMRG) \cite{White-DMRG-Article, White-DMRG-Letter, Schollwock-DMRG}. We call wave packets created using the full definition in Eq. \ref{eq:wave-packet} ``exact'', ``truncated'' wave packets are those whose sums are truncated to most dominant terms and use traditional fermionic creation operators, and ``truncated unitary// wave packets made with unitary creation operators (Fig. \ref{fig:Givens-rotation-circuit})y. For the truncated unitary wave packets, the ancilla sites are projected into $|0\rangle$-state prior to measurement. 

Figure \ref{fig:MPS-example} (a) shows the time evolved site occupations of a 16-site scattering simulation for $J=0.4,h=1,g=0.01$. The two wave packets are built in the subsystems of qubits 0-7 and 8-15 respectively. Figure \ref{fig:MPS-example} (c) shows the bipartite entanglement entropy that the regions spanned by sites 0 to $j+1$ have with the rest of the system for the truncated wave packets. The reason we don't use the truncated unitary packets is that the addition of control and ancilla qubits makes entanglement a non-trivial quantity to measure in the system (multipartite entanglement). Nevertheless, it should give a good idea of the entropy generated by the truncated unitary wave packets. Table \ref{tab:mps-results} shows the average relative error over the evolution of the occupation number and entanglement entropy between the exact and truncated wave packets and the average relative error in occupation between exact wave packets and the truncated unitary wave packets. For a given $J$, we get a good idea at the extent of perturbative $g$ regime and truncation. As you approach criticality, the perturbative window shrinks, but in most cases, the error is less than 10\%. We notice a slight deviation between the truncated and truncated unitary wave packets, which must come from a trace amount of leakage out of the system space. Nevertheless, we see that in this limit the truncated unitary wave packets are a good approximation to the exact wave packets.

\begin{table}
\centering
\begin{tabular}{|cc|cc|c|}
\hline
\multicolumn{2}{|c|}{Params.} & \multicolumn{2}{c|}{EWP-TWP \% Error}             & EWP-TUWP \% Error\\ \hline
\multicolumn{1}{|c|}{$J$} & $g$  & \multicolumn{1}{c|}{$\langle N_j(t) \rangle$} & $S_R(t)$ & $\langle N_j(t) \rangle$ \\ \hline
\multicolumn{1}{|c|}{0.4} & 0.01 & \multicolumn{1}{c|}{1.68}                     & 1.35     & 1.68                     \\ \hline
\multicolumn{1}{|c|}{}    & 0.02 & \multicolumn{1}{c|}{3.45}                     & 2.82     & 3.43                     \\ \hline
\multicolumn{1}{|c|}{}    & 0.03 & \multicolumn{1}{c|}{5.29}                     & 4.40     & 5.27                     \\ \hline
\multicolumn{1}{|c|}{}    & 0.05 & \multicolumn{1}{c|}{9.19}                     & 7.89     & 9.15                     \\ \hline
\multicolumn{1}{|c|}{0.6} & 0.01 & \multicolumn{1}{c|}{1.99}                     & 1.70     & 1.97                     \\ \hline
\multicolumn{1}{|c|}{}    & 0.02 & \multicolumn{1}{c|}{4.08}                     & 3.61     & 4.06                     \\ \hline
\multicolumn{1}{|c|}{}    & 0.03 & \multicolumn{1}{c|}{6.28}                     & 5.72     & 6.26                     \\ \hline
\multicolumn{1}{|c|}{}    & 0.05 & \multicolumn{1}{c|}{11.1}                     & 10.5     & 11.0                     \\ \hline
\multicolumn{1}{|c|}{0.8} & 0.01 & \multicolumn{1}{c|}{2.91}                     & 2.74     & 2.86                     \\ \hline
\multicolumn{1}{|c|}{}    & 0.02 & \multicolumn{1}{c|}{6.14}                     & 5.90     & 6.09                     \\ \hline
\multicolumn{1}{|c|}{}    & 0.03 & \multicolumn{1}{c|}{9.78}                     & 9.49     & 9.73                     \\ \hline
\multicolumn{1}{|c|}{1.0} & 0.01 & \multicolumn{1}{c|}{4.77}                     & 1.58     & 4.78                     \\ \hline
\multicolumn{1}{|c|}{}    & 0.02 & \multicolumn{1}{c|}{9.66}                     & 3.18     & 9.68                     \\ \hline
\multicolumn{1}{|c|}{}    & 0.03 & \multicolumn{1}{c|}{14.5}                     & 5.08     & 14.5                     \\ \hline
\end{tabular}
\caption{Average occupation and entanglement entropy relative percent errors over the entire time evolution (120 times steps) for a 16-site scattering state (see Fig. $\ref{fig:MPS-example}$) over a range of $J$ and $g$ ($h=1$). The second major column compares truncated wave packets (TWP) to exact wave packets (EWP). The third major column is the difference between exact wave packets and our truncated unitary wave packets (TUWP).}
\label{tab:mps-results}
\end{table}
\begin{figure}
    \centering
    \includegraphics[width=1.0\linewidth]{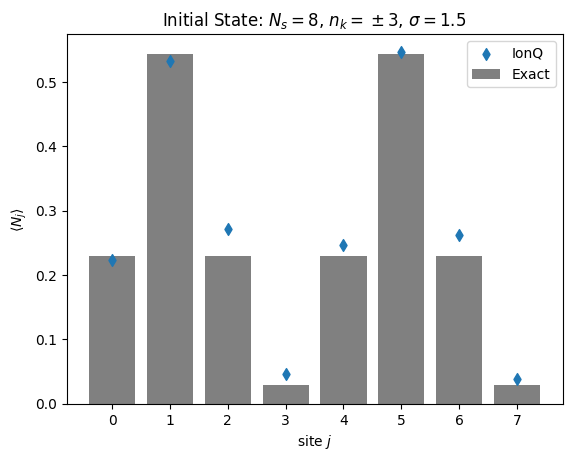}
    \caption{Initial state prepared on IonQ Forte 1 for an 8-site scattering state with initial positions $x=1,5$, momenta $k=\frac{3\pi}{8}$, and width $\sigma=3/2$. }
    \label{fig:ionq-run}
\end{figure}
For quantum simulation, we use VQE to construct the ground state with an efficient-SU(2) ansatz and Simultaneous Perturbation Stochastic Approximation (SPSA) optimizer \cite{qiskit2024}. For $N=8$ with $J=0.4, h=1, g=0.1$ and a single ansatz layer (CNOT depth $N-1$), the relative error in the energy of the VQE ground state and the true ground state is 0.53\%. Using IonQ's all-to-all connected Forte 1 processor, we prepare an initial scattering state on top of the VQE ground state. IonQ's built in debiasing was used for error correction \cite{ionq_docs}. Device data for the run is available here \cite{dataset}. Of the 5,000 total shots, 4,536 of those measured the ancilla's in the $|0\rangle$ state. Figure \ref{fig:ionq-run} shows the measured occupation of the IonQ initial state compared to an exact classically computed state. The relative percent error between the IonQ prepared state and the exact state is only 17.2\%. Neglecting the low occupation sites 3 and 7, the error is only 7.53\%. These results are promising for runs with limited error correction and sample size.
\begin{figure}
    \centering
    \includegraphics[width=1.0\linewidth]{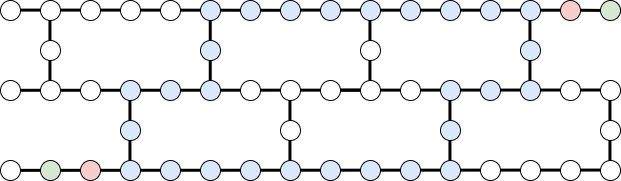}
    \caption{Example qubit map on IBM's heavy hexagonal topology. The blue circles represent system qubits, the white non-system qubits, and the red and green as control and ancilla qubits respectively that are used to implement $\sigma^-$. In this case, the system size is 28.}
    \label{fig:ibm-example}
\end{figure}

\textit{Conclusion} - Scattering simulations remain difficult in the NISQ era due to shallow circuit depths. We have shown that with minor adjustments to the original Givens Rotation method for fermionic scattering state preparation \cite{chai2024fermionicwavepacketscattering}, we have provided an accurate and efficient method to generate scattering states for weakly interacting theories both near and at criticality. This was done by localizing the wave packets in space allowing for parallel CNOT execution, and adding ladder operator block encoding \cite{Simon2025ladderoperatorblock} for true unitary particle creation operators. With the method containing at most nearest neighbor two and three qubit gates, it can be readily implemented on transmon machines like IBM (Figure \ref{fig:ibm-example}). Via MPS simulations, we verified that these localized wave packets are a good approximation for weakly interacting theories both at and away from criticality. Using IonQ's Forte 1 machine, we showed that our methods can produce accurate initial states with today's hardware. Future work will include simulating strongly interacting theories which involves overlapping the individual wave packet regions at the cost of increased CNOT depth, explore staggered theories like Gross-Neveu, and to compare with variational approaches. Thus, we have an algorithm for the near term and future of quantum scattering simulations.

\textit{Acknowledgments} - This work is supported in part by the Department of Energy under Award Numbers DE-SC0019139 and DE-SC0010113 and NSF award DMR-1747426. I would like to thank Yannick Meurice, Muhammad Asaduzzaman, Cameron Cogburn and Zheyue Hang for helpful discussions. I also thank Nikita Zemlevskiy and Zane Ozzello for comments on the original manuscript.

\textit{Data Availability} - The data that support the findings of this article are openly available \cite{dataset}.
\bibliography{ref}

@data{dataset,
author = {Hite, Michael},
publisher = {Harvard Dataverse},
title = {{Replication Data for: Improved Fermionic Scattering for the NISQ Era}},
year = {2025},
version = {DRAFT VERSION},
doi = {10.7910/DVN/NGAY3D},
url = {https://doi.org/10.7910/DVN/NGAY3D}
}

@misc{ionq_docs,
  author = {{IonQ, Inc.}},
  title = {IonQ Quantum Cloud Documentation},
  year = {2026},
  url = {https://docs.ionq.com/},
  note = {Accessed: 2026-02-11}
}

@book{Sachdev, 
    place={Cambridge}, 
    edition={2}, 
    title={Quantum Phase Transitions}, 
    publisher={Cambridge University Press}, 
    author={Sachdev, Subir}, 
    year={2011}
}

@article{Kaufman,
    title = {Crystal Statistics. II. Partition Function Evaluated by Spinor Analysis},
    author = {Kaufman, Bruria},
    journal = {Phys. Rev.},
    volume = {76},
    issue = {8},
    pages = {1232--1243},
    numpages = {0},
    year = {1949},
    month = {Oct},
    publisher = {American Physical Society},
    doi = {10.1103/PhysRev.76.1232},
    url = {https://link.aps.org/doi/10.1103/PhysRev.76.1232}
}

@article{SML,
    title = {Two-Dimensional Ising Model as a Soluble Problem of Many Fermions},
    author = {Schultz, T. D. and Mattis, D. C. and Lieb, E. H.},
    journal = {Rev. Mod. Phys.},
    volume = {36},
    issue = {3},
    pages = {856--871},
    numpages = {0},
    year = {1964},
    month = {Jul},
    publisher = {American Physical Society},
    doi = {10.1103/RevModPhys.36.856},
    url = {https://link.aps.org/doi/10.1103/RevModPhys.36.856}
}

@article{lie-trotter,
 ISSN = {00029939, 10886826},
 URL = {http://www.jstor.org/stable/2033649},
 author = {H. F. Trotter},
 journal = {Proceedings of the American Mathematical Society},
 number = {4},
 pages = {545--551},
 publisher = {American Mathematical Society},
 title = {On the Product of Semi-Groups of Operators},
 urldate = {2025-12-28},
 volume = {10},
 year = {1959}
}

@article{suzuki-trotter,
author = {Masuo Suzuki},
title = {{Generalized Trotter's formula and systematic approximants of exponential operators and inner derivations with applications to many-body problems}},
volume = {51},
journal = {Communications in Mathematical Physics},
number = {2},
publisher = {Springer},
pages = {183 -- 190},
year = {1976},
}

@misc{qiskit2024,
    title={Quantum computing with {Q}iskit},
    author={Javadi-Abhari, Ali and Treinish, Matthew and Krsulich, Kevin and Wood, Christopher J. and Lishman, Jake and Gacon, Julien and Martiel, Simon and Nation, Paul D. and Bishop, Lev S. and Cross, Andrew W. and Johnson, Blake R. and Gambetta, Jay M.},
    year={2024},
    doi={10.48550/arXiv.2405.08810},
    eprint={2405.08810},
    archivePrefix={arXiv},
    primaryClass={quant-ph}
}

@Article{ITensorArticle,
	title={{The ITensor Software Library for Tensor Network Calculations}},
	author={Matthew Fishman and Steven R. White and E. Miles Stoudenmire},
	journal={SciPost Phys. Codebases},
	pages={4},
	year={2022},
	publisher={SciPost},
	doi={10.21468/SciPostPhysCodeb.4},
	url={https://scipost.org/10.21468/SciPostPhysCodeb.4},
}

@Article{ITensorCode,
	title={{Codebase release 0.3 for ITensor}},
	author={Matthew Fishman and Steven R. White and E. Miles Stoudenmire},
	journal={SciPost Phys. Codebases},
	pages={4-r0.3},
	year={2022},
	publisher={SciPost},
	doi={10.21468/SciPostPhysCodeb.4-r0.3},
	url={https://scipost.org/10.21468/SciPostPhysCodeb.4-r0.3},
}

@article{Preskill2018quantumcomputingin,
  doi = {10.22331/q-2018-08-06-79},
  url = {https://doi.org/10.22331/q-2018-08-06-79},
  title = {Quantum {C}omputing in the {NISQ} era and beyond},
  author = {Preskill, John},
  journal = {{Quantum}},
  issn = {2521-327X},
  publisher = {{Verein zur F{\"{o}}rderung des Open Access Publizierens in den Quantenwissenschaften}},
  volume = {2},
  pages = {79},
  month = aug,
  year = {2018}
}

@article{Simon2025ladderoperatorblock,
  doi = {10.22331/q-2025-12-22-1953},
  url = {https://doi.org/10.22331/q-2025-12-22-1953},
  title = {Ladder {O}perator {B}lock-{E}ncoding},
  author = {Simon, William A. and Gustin, Carter M. and Serafin, Kamil and Ralli, Alexis and Goldstein, Gary R. and Love, Peter J.},
  journal = {{Quantum}},
  issn = {2521-327X},
  publisher = {{Verein zur F{\"{o}}rderung des Open Access Publizierens in den Quantenwissenschaften}},
  volume = {9},
  pages = {1953},
  month = dec,
  year = {2025}
}

@misc{mottonen2004transformationquantumstatesusing,
    title={Transformation of quantum states using uniformly controlled rotations}, 
    author={Mikko Mottonen and Juha J. Vartiainen and Ville Bergholm and Martti M. Salomaa},
    year={2004},
    eprint={quant-ph/0407010},
    archivePrefix={arXiv},
    primaryClass={quant-ph},
    url={https://arxiv.org/abs/quant-ph/0407010}
}

@article{Shende_2006,
    title={Synthesis of quantum-logic circuits},
    volume={25},
    ISSN={1937-4151},
    url={http://dx.doi.org/10.1109/TCAD.2005.855930},
    DOI={10.1109/tcad.2005.855930},
    number={6},
    journal={IEEE Transactions on Computer-Aided Design of Integrated Circuits and Systems},
    publisher={Institute of Electrical and Electronics Engineers (IEEE)},
    author={Shende, V.V. and Bullock, S.S. and Markov, I.L.},
    year={2006},
    month=jun, pages={1000–1010}
}

@misc{Arrazola_2021,
    title = "Givens rotations for quantum chemistry",
    author = "Juan Miguel Arrazola",
    year = "2021",
    month = "06",
    journal = "PennyLane Demos",
    publisher = "Xanadu",
    howpublished = {https://pennylane.ai/qml/demos/tutorial\_givens\_rotations}
}

@article{Arrazola_2022,
    title={Universal quantum circuits for quantum chemistry},
    volume={6},
    ISSN={2521-327X},
    url={http://dx.doi.org/10.22331/q-2022-06-20-742},
    DOI={10.22331/q-2022-06-20-742},
    journal={Quantum},
    publisher={Verein zur Forderung des Open Access Publizierens in den Quantenwissenschaften},
    author={Arrazola, Juan Miguel and Di Matteo, Olivia and Quesada, Nicolás and Jahangiri, Soran and Delgado, Alain and Killoran, Nathan},
    year={2022},
    month=jun, pages={742}
}

@article{PhysRevA.98.022322,
    title = {Quantum algorithms for electronic structure calculations: Particle-hole Hamiltonian and optimized wave-function expansions},
    author = {Barkoutsos, Panagiotis Kl. and Gonthier, Jerome F. and Sokolov, Igor and Moll, Nikolaj and Salis, Gian and Fuhrer, Andreas and Ganzhorn, Marc and Egger, Daniel J. and Troyer, Matthias and Mezzacapo, Antonio and Filipp, Stefan and Tavernelli, Ivano},
    journal = {Phys. Rev. A},
    volume = {98},
    issue = {2},
    pages = {022322},
    numpages = {13},
    year = {2018},
    month = {Aug},
    publisher = {American Physical Society},
    doi = {10.1103/PhysRevA.98.022322},
    url = {https://link.aps.org/doi/10.1103/PhysRevA.98.022322}
}

@article{Peruzzo2014,
    author={Peruzzo, Alberto and McClean, Jarrod and Shadbolt, Peter and Yung, Man-Hong and Zhou, Xiao-Qi and Love, Peter J. and Aspuru-Guzik, Alan and O'Brien, Jeremy L.},
    title={A variational eigenvalue solver on a photonic quantum processor},
    journal={Nature Communications},
    year={2014},
    month={Jul},
    day={23},
    volume={5},
    number={1},
    pages={4213},
    issn={2041-1723},
    doi={10.1038/ncomms5213},
    url={https://doi.org/10.1038/ncomms5213}
}

@misc{shende2008cnotcosttoffoligates,
      title={On the CNOT-cost of TOFFOLI gates}, 
      author={Vivek V. Shende and Igor L. Markov},
      year={2008},
      eprint={0803.2316},
      archivePrefix={arXiv},
      primaryClass={quant-ph},
      url={https://arxiv.org/abs/0803.2316}, 
}

@article{Jordan_2012,
   title={Quantum Algorithms for Quantum Field Theories},
   volume={336},
   ISSN={1095-9203},
   url={http://dx.doi.org/10.1126/science.1217069},
   DOI={10.1126/science.1217069},
   number={6085},
   journal={Science},
   publisher={American Association for the Advancement of Science (AAAS)},
   author={Jordan, Stephen P. and Lee, Keith S. M. and Preskill, John},
   year={2012},
   month=jun, pages={1130–1133}
}

@misc{jordan2014quantumalgorithmsfermionicquantum,
      title={Quantum Algorithms for Fermionic Quantum Field Theories}, 
      author={Stephen P. Jordan and Keith S. M. Lee and John Preskill},
      year={2014},
      eprint={1404.7115},
      archivePrefix={arXiv},
      primaryClass={hep-th},
      url={https://arxiv.org/abs/1404.7115}, 
}

@article{qsimscatt,
  title = {Quantum simulation of scattering in the quantum Ising model},
  author = {Gustafson, Erik and Meurice, Y. and Unmuth-Yockey, Judah},
  journal = {Phys. Rev. D},
  volume = {99},
  issue = {9},
  pages = {094503},
  numpages = {13},
  year = {2019},
  month = {May},
  publisher = {American Physical Society},
  doi = {10.1103/PhysRevD.99.094503},
  url = {https://link.aps.org/doi/10.1103/PhysRevD.99.094503}
}

@article{OG-PhaseShiftPaper,
  title = {Real-time quantum calculations of phase shifts using wave packet time delays},
  author = {Gustafson, Erik and Zhu, Yingyue and Dreher, Patrick and Linke, Norbert M. and Meurice, Yannick},
  journal = {Phys. Rev. D},
  volume = {104},
  issue = {5},
  pages = {054507},
  numpages = {11},
  year = {2021},
  month = {Sep},
  publisher = {American Physical Society},
  doi = {10.1103/PhysRevD.104.054507},
  url = {https://link.aps.org/doi/10.1103/PhysRevD.104.054507}
}

@misc{parks2023applyingnoxerrormitigation,
      title={Applying NOX Error Mitigation Protocols to Calculate Real-time Quantum Field Theory Scattering Phase Shifts}, 
      author={Zachary Parks and Arnaud Carignan-Dugas and Erik Gustafson and Yannick Meurice and Patrick Dreher},
      year={2023},
      eprint={2212.05333},
      archivePrefix={arXiv},
      primaryClass={quant-ph},
      url={https://arxiv.org/abs/2212.05333}, 
}

@misc{chai2024fermionicwavepacketscattering,
      title={Fermionic wave packet scattering: a quantum computing approach}, 
      author={Yahui Chai and Arianna Crippa and Karl Jansen and Stefan Kühn and Vincent R. Pascuzzi and Francesco Tacchino and Ivano Tavernelli},
      year={2024},
      eprint={2312.02272},
      archivePrefix={arXiv},
      primaryClass={quant-ph},
      url={https://arxiv.org/abs/2312.02272}, 
}

@misc{davoudi2024scatteringwavepacketshadrons,
      title={Scattering wave packets of hadrons in gauge theories: Preparation on a quantum computer}, 
      author={Zohreh Davoudi and Chung-Chun Hsieh and Saurabh V. Kadam},
      year={2024},
      eprint={2402.00840},
      archivePrefix={arXiv},
      primaryClass={quant-ph},
      doi={https://doi.org/10.22331/q-2024-11-11-1520},
      url={https://arxiv.org/abs/2402.00840}, 
}

@misc{farrell2024quantumsimulationshadrondynamics,
      title={Quantum Simulations of Hadron Dynamics in the Schwinger Model using 112 Qubits}, 
      author={Roland C. Farrell and Marc Illa and Anthony N. Ciavarella and Martin J. Savage},
      year={2024},
      eprint={2401.08044},
      archivePrefix={arXiv},
      primaryClass={quant-ph},
      url={https://arxiv.org/abs/2401.08044}, 
}

@misc{Zemlevskiy:2024vxt,
      title={Scalable Quantum Simulations of Scattering in Scalar Field Theory on 120 Qubits}, 
      author={Nikita A. Zemlevskiy},
      year={2024},
      eprint={2411.02486},
      archivePrefix={arXiv},
      primaryClass={quant-ph},
      url={https://arxiv.org/abs/2411.02486}, 
}

@article{Turco_2024,
   title={Quantum Simulation of Bound State Scattering},
   volume={5},
   ISSN={2691-3399},
   url={http://dx.doi.org/10.1103/PRXQuantum.5.020311},
   DOI={10.1103/prxquantum.5.020311},
   number={2},
   journal={PRX Quantum},
   publisher={American Physical Society (APS)},
   author={Turco, Matteo and Quinta, Gonçalo and Seixas, João and Omar, Yasser},
   year={2024},
   month={apr}
}

@misc{kreshchuk2023simulatingscatteringcompositeparticles,
      title={Simulating Scattering of Composite Particles}, 
      author={Michael Kreshchuk and James P. Vary and Peter J. Love},
      year={2023},
      eprint={2310.13742},
      archivePrefix={arXiv},
      primaryClass={quant-ph},
      url={https://arxiv.org/abs/2310.13742}, 
}

@misc{briceño2023coherentquantumcomputationscattering,
      title={Toward coherent quantum computation of scattering amplitudes with a measurement-based photonic quantum processor}, 
      author={Raúl A. Briceño and Robert G. Edwards and Miller Eaton and Carlos González-Arciniegas and Olivier Pfister and George Siopsis},
      year={2023},
      eprint={2312.12613},
      archivePrefix={arXiv},
      primaryClass={quant-ph},
      url={https://arxiv.org/abs/2312.12613}, 
}

@article{Sharma_2024,
   title={Scattering phase shifts from a quantum computer},
   volume={109},
   ISSN={2469-9993},
   url={http://dx.doi.org/10.1103/PhysRevC.109.L061001},
   DOI={10.1103/physrevc.109.l061001},
   number={6},
   journal={Physical Review C},
   publisher={American Physical Society (APS)},
   author={Sharma, Sanket and Papenbrock, T. and Platter, L.},
   year={2024},
   month={jun}
}

@article{Wang_2024,
   title={Nuclear scattering via quantum computing},
   volume={109},
   ISSN={2469-9993},
   url={http://dx.doi.org/10.1103/PhysRevC.109.064623},
   DOI={10.1103/physrevc.109.064623},
   number={6},
   journal={Physical Review C},
   publisher={American Physical Society (APS)},
   author={Wang, Peiyan and Du, Weijie and Zuo, Wei and Vary, James P.},
   year={2024},
   month={jun}
}

@misc{bennewitz2024simulatingmesonscatteringspin,
      title={Simulating Meson Scattering on Spin Quantum Simulators}, 
      author={Elizabeth R. Bennewitz and Brayden Ware and Alexander Schuckert and Alessio Lerose and Federica M. Surace and Ron Belyansky and William Morong and De Luo and Arinjoy De and Kate S. Collins and Or Katz and Christopher Monroe and Zohreh Davoudi and Alexey V. Gorshkov},
      year={2024},
      eprint={2403.07061},
      archivePrefix={arXiv},
      primaryClass={quant-ph},
      url={https://arxiv.org/abs/2403.07061}, 
}

@misc{turro2024evaluationphaseshiftsnonrelativistic,
      title={Evaluation of phase shifts for non-relativistic elastic scattering using quantum computers}, 
      author={Francesco Turro and Kyle A. Wendt and Sofia Quaglioni and Francesco Pederiva and Alessandro Roggero},
      year={2024},
      eprint={2407.04155},
      archivePrefix={arXiv},
      primaryClass={quant-ph},
      url={https://arxiv.org/abs/2407.04155}, 
}

@misc{yusf2024elasticscatteringquantumcomputer,
      title={Elastic scattering on a quantum computer}, 
      author={Muhammad Yusf and Ling Gan and Cameron Moffat and Gautam Rupak},
      year={2024},
      eprint={2406.09231},
      archivePrefix={arXiv},
      primaryClass={nucl-th},
      url={https://arxiv.org/abs/2406.09231}, 
}

@misc{farrell2025digitalquantumsimulationsscattering,
      title={Digital quantum simulations of scattering in quantum field theories using W states}, 
      author={Roland C. Farrell and Nikita A. Zemlevskiy and Marc Illa and John Preskill},
      year={2025},
      eprint={2505.03111},
      archivePrefix={arXiv},
      primaryClass={quant-ph},
      url={https://arxiv.org/abs/2505.03111}, 
}

@article{Cervera_Lierta_2018,
   title={Exact Ising model simulation on a quantum computer},
   volume={2},
   ISSN={2521-327X},
   url={http://dx.doi.org/10.22331/q-2018-12-21-114},
   DOI={10.22331/q-2018-12-21-114},
   journal={Quantum},
   publisher={Verein zur Forderung des Open Access Publizierens in den Quantenwissenschaften},
   author={Cervera-Lierta, Alba},
   year={2018},
   month=dec, pages={114}
}

@article{White-DMRG-Letter,
    title = {Density matrix formulation for quantum renormalization groups},
    author = {White, Steven R.},
    journal = {Phys. Rev. Lett.},
    volume = {69},
    issue = {19},
    pages = {2863--2866},
    numpages = {0},
    year = {1992},
    month = {Nov},
    publisher = {American Physical Society},
    doi = {10.1103/PhysRevLett.69.2863},
    url = {https://link.aps.org/doi/10.1103/PhysRevLett.69.2863}
}

@article{White-DMRG-Article,
    title = {Density-matrix algorithms for quantum renormalization groups},
    author = {White, Steven R.},
    journal = {Phys. Rev. B},
    volume = {48},
    issue = {14},
    pages = {10345--10356},
    numpages = {0},
    year = {1993},
    month = {Oct},
    publisher = {American Physical Society},
    doi = {10.1103/PhysRevB.48.10345},
    url = {https://link.aps.org/doi/10.1103/PhysRevB.48.10345}
}

@article{Schollwock-DMRG,
    title = {The density-matrix renormalization group},
    author = {Schollw\"ock, U.},
    journal = {Rev. Mod. Phys.},
    volume = {77},
    issue = {1},
    pages = {259--315},
    numpages = {0},
    year = {2005},
    month = {Apr},
    publisher = {American Physical Society},
    doi = {10.1103/RevModPhys.77.259},
    url = {https://link.aps.org/doi/10.1103/RevModPhys.77.259}
}

@article{TEBD-OG-Paper,
    title = {Efficient Classical Simulation of Slightly Entangled Quantum Computations},
    author = {Vidal, Guifr\'e},
    journal = {Phys. Rev. Lett.},
    volume = {91},
    issue = {14},
    pages = {147902},
    numpages = {4},
    year = {2003},
    month = {Oct},
    publisher = {American Physical Society},
    doi = {10.1103/PhysRevLett.91.147902},
    url = {https://link.aps.org/doi/10.1103/PhysRevLett.91.147902}
}

\end{document}